\documentclass[aps,prb,twocolumn,superscriptaddress,showpacs]{revtex4}
\usepackage{graphicx}
\usepackage{latexsym}
\usepackage{amssymb}
\usepackage{amsmath}
\usepackage{amsfonts}
\usepackage{bm}
\usepackage{multirow}
\usepackage{color}
\usepackage{comment}
\newcommand{\ii}{\mathrm{i}}

\newcommand{\ve}{\vec{e}}

\newcommand{\va}{\vec{a}}

\newcommand{\SO}{\mathrm{SO}}

\newcommand{\SU}{\mathrm{SU}}
\newcommand{\U}{\mathrm{U}}

\newcommand{\beq}{\begin{equation}}
\newcommand{\eeq}{\end{equation}}
\newcommand{\beqn}{\begin{eqnarray}}
\newcommand{\eeqn}{\end{eqnarray}}

\DeclareMathAlphabet{\mathbbold}{U}{bbold}{m}{n}

\def\SU{{\rm SU}}

\def\U{{\rm U}}

\begin{document}

\title{Physics of Symmetry Protected Topological phases involving Higher Symmetries and their Applications}


\author{Chao-Ming Jian}
\affiliation{Department of Physics, Cornell University, Ithaca,
New York 14853, USA}

\author{Xiao-Chuan Wu}
\affiliation{Department of Physics, University of California,
Santa Barbara, CA 93106, USA}

\author{Yichen Xu}
\affiliation{Department of Physics, University of California,
Santa Barbara, CA 93106, USA}

\author{Cenke Xu}
\affiliation{Department of Physics, University of California,
Santa Barbara, CA 93106, USA}

\begin{abstract}

We discuss physical constructions, and the boundary properties of
various symmetry protected topological phases that involve 1-form
symmetries, from one spatial dimension ($1d$) to four spatial
dimensions ($4d$). For example, the prototype $3d$ boundary state
of $4d$ SPT states involving 1-form symmetries can be either a
gapless photon phase (quantum electrodynamics) or gapped
topological order enriched by 1-form symmetries, namely the loop
excitations of these topological orders carry nontrivial 1-form
symmetry charges. This study also serves the purpose of diagnosing
anomaly of $3d$ states of matter.
Connection between SPT states with 1-form symmetries and condensed
matter systems such as quantum dimer models at one lower dimension
will also be discussed. Whether a quantum dimer model can have a
trivial gapped phase or not depends on the nature of its
corresponding bulk state in one higher dimension.

\end{abstract}

\maketitle

\section{Introduction}

The symmetry protected topological (SPT)
phases~\cite{wenspt,wenspt2} have greatly enriched our
understanding of quantum states of matter. With certain
symmetries, the boundary of these SPT states cannot be trivially
gapped without degeneracy. Especially, many exotic states of
matter can be realized at the $2d$ boundary of $3d$ bosonic SPT
states. For example, exotic quantum critical points (QCP) in $2d$
with spatial symmetries (both on the square or triangular lattice)
can be realized at the boundary of certain $3d$ SPT
states~\cite{senthilashvin,xutriangle}, and the conjectured
emergent symmetry of the deconfined QCP matches well with the bulk
symmetry of the SPT state, sometimes these emergent symmetries are
only revealed through certain dualities~\cite{xudual1,xudual2}
between $(2+1)d$ quantum field theories. The analysis of the SPT
state in the $(d+1)$-dimensional bulk can also be used as a
diagnose of the ``Lieb-Schultz-Mattis theorem" in $d$-dimensional
systems with spatial symmetries, i.e. whether or not the
$d$-dimensional system can be gapped without
degeneracy~\cite{xulsm,cholsm,maxlsm,chenglsm,masakilsm,elselsm}
is related to the nature of the corresponding bulk state in one
higher dimension.

In recent years it was realized that the very concept of symmetry
can be generalized to higher dimensional objects rather than just
point like
operators~\cite{formsym1,formsym2,formsym3,formsym4,formsym5,formsym6,formsym7,formsym8,Cordova2019}.
Examples of SPT states that involve these generalized symmetries
were discussed in previous
literatures~\cite{mcgreevy,formsym6,thorngren2015higher,zhu2019,Cordova2019,xubsre,ye2014,juven,juven2,wen1formspt1,wen1formspt2}.
For example a classification of SPT states based on generalized
cobordism theory was given in Ref.~\onlinecite{juven,juven2},
exactly soluble lattice models for a class of SPT states were
constructed in Ref.~\onlinecite{wen1formspt1,wen1formspt2}. In the
current manuscript we focus on physical construction and boundary
properties of a series of SPT states with generalized concepts of
symmetries, from $(1+1)$-dimension to $(4+1)$-dimension. We do not
seek for exactly soluble models, instead we will focus on general
physical pictures of these states. For example, the prototype $4d$
(or $(4+1)d$) SPT state we will discuss can be constructed by
``decorated Dirac monopole loop" picture, which is analogous to
the flux attachment construction in $2d$ SPT state. And the
prototype $3d$ boundary state of the $4d$ SPT state is a photon
phase with various constraints of dynamics, quantum numbers, and
statistics on the electric and magnetic charges. We assume that
the gauge invariant objects/excitations, i.e. objects that do not
couple to {\it dynamical} gauge field, are always bosonic. These
include point particles and higher dimensional excitations such as
loops.

The 1-form symmetry transformation acts on loop-like operators
such as the Wilson loop or 't Hooft loop of a dynamical gauge
field. The existence of an electric 1-form symmetry demands that
the electric charge of the gauge field is infinitely heavy. In
condensed matter systems the quantum dimer model~\cite{rk}
naturally fits this criterion. It is well-known that the quantum
dimer model can be mapped to a lattice gauge field~\cite{qd}. In a
quantum dimer model, every site of the lattice is connected to a
fixed number of dimers, which implies that there is a background
electric charge distribution, but no dynamical charge in the
system. Hence the quantum dimer model naturally has a 1-form
symmetry. The quantum dimer model on certain $d$-dimensional
lattice may be mapped to the boundary of a $(d+1)$-dimensional SPT
state with 1-form symmetry in certain limit, and the spatial
symmetries of the quantum dimer model is mapped to the onsite
symmetry of the bulk SPT state. The analysis of the SPT state in
the bulk has strong indications on the allowed phenomena of the
quantum dimer model at $d$-dimension.

Due to the inevitable complexity of notations used in this
manuscript, we will keep a self-consistent conventions of
notations:

The $N-$form symmetry $G$ will be labelled as $G^{(N)}$, such as
$\U(1)^{(1)}$, $Z_n^{(1)}$, etc. Ordinary 0-form symmetry will be
labelled without superscript.

Gauge symmetries associated with {\it dynamical} gauge field will
be labelled as $u(1)^{(1)}$, $z_n^{(2)}$, etc. depending on the
nature of the gauge fields. A topological order which corresponds
to a dynamical discrete gauge field will also be labelled as, for
example, a $z_n$ topological order.

Gauge symmetries associated with {\it background} gauge fields
will be labelled as $\mathcal{U}(1)^{(1)}$, $\mathcal{Z}_n^{(2)}$,
etc.

Classifications of SPT states will be labelled as $\mathbb{Z}$,
$\mathbb{Z}_n$, etc.

For space and space-time dimensions, for example, $3d$ space
refers to three spatial dimensions; $(3+1)d$ refers to the
space-time dimension, which is the same as $4D$ Euclidean
space-time. Also, QED$_4$ refers to quantum electrodynamics in
$(3+1)d$ or $4D$ space-time dimension.

For a QED$_4$, there are point like particles such as electric
charge, and Dirac monopole. We label bosonic (fermionic) electric
charges as $e_b$ $(e_f)$, and bosonic (fermionic) Dirac monopoles
as $m_b$ $(m_f)$. Some of these point excitations have no dynamics
(infinitely heavy) due to the 1-form symmetries, we will label
these immobile point particles as $e_{0b}$, $e_{0f}$, etc. A
QED$_4$ with bosonic electric charge and fermionic Dirac monopole
is labelled as ``$\mathrm{QED}_4\{e_b, m_f\}$".

\section{Building bricks: $1d$ SPT state with 1-form symmetries}

\label{1d}

The simplest SPT state that involves a 1-form symmetry exists in
$1d$ space or $(1+1)d$ space-time. $1d$ SPT state with a 1-form
symmetry is analogous to an ordinary SPT state in $0d$ space. For
a $\U(1)^{(1)}$ 1-form symmetry, a SPT state in $1d$ simply
corresponds to a state with integer electric flux through the
system. Let us take a $1d$ chain with electric field operators
defined on the links. Due to the Gauss law constraint, $\nabla_x
\hat{e}(x) = 0$, the electric field $\hat{e}(x)$ takes a uniform
integer eigenvalue on the entire chain (in a compact $u(1)$
lattice gauge theory, the electric field operator $\hat{e}(x)$
takes discrete integer value, while its conjugate operator
$\hat{a}(x)$ is periodically defined), hence for a $\U(1)^{(1)}$
1-form symmetry, the classification of $1d$ SPT states is
$\mathbb{Z}$, which corresponds to different integer eigenvalues
of $\hat{e}(x)$. It is analogous to the $\mathbb{Z}$
classification of a zero dimensional ordinary SPT state with
$\U(1)$ symmetry~\cite{wenspt,wenspt2}.

The Hamiltonian of a $1d$ lattice $\U(1)$ gauge field is also very
simple, for example: \beqn H = \sum_x \ g \left( \hat{e}(x) - k
\right)^2. \label{1dh} \eeqn Due to the Gauss law constraint, a
Hamiltonian must be invariant under gauge transformation $\hat{a}
\rightarrow \hat{a} + \nabla_x f(x)$, where $\hat{a}$ is the
conjugate operator of $\hat{e}$. A local $1d$ Hamiltonian that
involves $\hat{a}$ cannot be gauge invariant, hence a local gauge
invariant Hamiltonian is only a function of $\hat{e}$. In
Eq.~\ref{1dh} $k$ can take continuous values. When $k$ is half
integer, the system is at the transition between two SPT states,
and the ground state of the Hamiltonian is two-fold degenerate
with $\hat{e}(x) = k \pm 1/2$, namely the transition is a level
crossing between two eigenvalues of $\hat{e}(x)$. This transition
should be viewed as a first order transition.

One can also couple the electric field to a background 2-form
$\mathcal{U}(1)^{(2)}$ gauge field: \beqn S = \int d\tau dx \ \ii
f_{\mu\nu} B_{\mu\nu} \eeqn In $(1+1)d$ the stress tensor of the
$u(1)$ gauge field is just the electric field: $f_{10} - f_{01} =
e(x)$, and $B_{01} = - B_{10}$ is a Lagrange multiplier. Hence the
$(1+1)d$ topological response theory for the SPT state is \beqn
S_{\mathrm{1d-topo}} = \int_{(1+1)d} \ \ii k B,
\label{1dtopo}\eeqn which is a $(1+1)d$ Chern-Simons action of the
2-form gauge field $B$, and its level $k$ takes only integer
values. For each integer level$-k$, the electric field (the 1-form
symmetry charge) \beqn e(x) = \frac{\delta S_{\mathrm{1d-topo}}}
{\ii \delta B(x)} = k. \eeqn

The $1d$ SPT state with 1-form symmetries will be the building
bricks for SPT states in higher dimensions. Suppose we break the
$\U(1)^{(1)}$ down to $Z_n^{(1)}$ symmetry, the topological
response theory Eq.~\ref{1dtopo} still applies, but $B$ is now a
2-form $\mathcal{Z}_n^{(2)}$ background gauge field. The
classification of the SPT state will reduce to $\mathbb{Z}_n$,
which means that in Eq.~\ref{1dtopo} the integer $k + n = k$.

\section{$4d$ SPT states with $G_1^{(1)}\times
G_2^{(1)}$ symmetry}

\label{4d1}

\subsection{Parent $4d$ SPT state with $\U(1)^{(1)}\times
\U(1)^{(1)}$ symmetry}

We now discuss SPT states in $4d$ space that involves 1-form
symmetries. This discussion is useful for diagnosing anomalies of
$3d$ states of matter, namely some $3d$ states of matter can only
be realized at the boundary of a $4d$ SPT state. The parent SPT
state that we will start with is the $(4+1)d$ state with the
$\U(1)^{(1)}\times \U(1)^{(1)}$ 1-form symmetry. With two
$\U(1)^{(1)}$ 1-form symmetries, the system can couple to two
background $\mathcal{U}(1)^{(2)}$ 2-form gauge fields $B^1$ and
$B^2$, and the response theory in $(4+1)d$ reads \beqn
S_{\mathrm{4d-topo}} = \int_{(4+1)d} \ \frac{\ii k}{4\pi}
\epsilon_{IJ} B^I \wedge d B^J, \label{4dtopo} \eeqn where
$\epsilon_{IJ} = \ii \sigma^y$. For each integer $k$,
Eq.~\ref{4dtopo} is a different Chern-Simons theory, and the
system should correspond to a different SPT state, hence these SPT
states described by Eq.~\ref{4dtopo} have a $\mathbb{Z}$
classification. The $(3+1)d$ boundary of this state is a QED$_4$
without dynamical electric or magnetic charge (Dirac monopole).
This QED$_4$ has a $\U(1)^{(1)}\times \U(1)^{(1)}$ mixed 't Hooft
anomaly as was derived in previous
literatures~\cite{mcgreevy,formsym6,Cordova2019}.

To construct this $4d$ SPT state, we can start with two $(4+1)d$
$u(1)$ gauge fields $\va^1$ and $\va^{2}$. These two gauge fields
both have {\it electric} 1-form $\U(1)^{(1)}$ symmetry, namely
both gauge fields have no dynamical electric charges, i.e. the
Gauss law constraint on the electric field is strictly enforced.
This is equivalent to tuning the electric charges in the $4d$ bulk
to be infinitely heavy. Both $u(1)$ gauge fields allow dynamical
Dirac monopole loop/line defects in the $4d$ space.
We will first discuss the cases where the charges of $\va^1$ and
$\va^2$ are both bosons, otherwise $\va^1$ and $\va^2$ would be
Spin$^C$ connections.
Situations with fermionic gauge charges of $\va^1$ and $\va^2$
will be discussed later.

\begin{figure}
\includegraphics[width=0.38\textwidth]{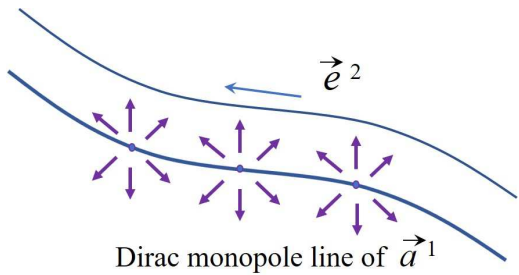}
\caption{The decorated Dirac monopole loop construction of the
parent SPT state in $4d$ space. The Dirac monopole loop of gauge
field $\vec{a}^1$ is decorated with the $1d$ SPT state of the
$\U(1)^{(1)}$ 1-form symmetry associated with gauge field
$\vec{a}^2$. After the condensation of the decorated Dirac
monopole loops, the $4d$ system is driven into a SPT state
described by response theory Eq.~\ref{4dtopo}. } \label{decorate}
\end{figure}

We use the analogue of the ``flux attachment" (or ``decorated
defect") construction of the SPT state which was used to construct
$2d$ bosonic SPT state~\cite{levinsenthil}. In $2d$ space, a
$\U(1) \times \U(1)$ SPT state (the parent state of many $2d$ SPT
states) can be constructed by binding the vortex defect of one
$\U(1)$ symmetry with the charge of the other $\U(1)$ symmetry,
and condense the bound state, which drives the system into a
gapped SPT phase. In $4d$ space, the analogue of the vortex defect
of an ordinary $\U(1)$ 0-form symmetry, is the Dirac monopole
loop/line of a $u(1)$ gauge field. We decorate the Dirac monopole
loop of $\va^1$ with the $1d$ SPT state defined with the 1-form
symmetry associated with $\va^2$ with level $(+k)$ in
Eq.~\ref{1dtopo}, and condense/proliferate the decorated loops
(Fig.~\ref{decorate}).
Once the bound state between the monopole loop of $\va^1$ and the
$(+k)$ unit of electric flux of $\va^2$ is condensed, the monopole
loop of $\va^2$ will be automatically bound with $(-k)$ unit of
electric flux of $\va^1$.

Condensation of Dirac monopole loops would normally drive a
$(4+1)d$ $u(1)$ gauge field to the gapped confined phase (the loop
excitation is coupled to a dual dynamical 2-form gauge field, and
the condensate is gapped due to the Higgs mechanism). But because
the Dirac monopole loop is decorated with another SPT state with
$1-$form symmetry in our case, after the condensation of the
decorated monopole loops, the phase in the $4d$ bulk is not an
ordinary confined phase, it is actually a SPT phase described by
Eq.~\ref{4dtopo}. In fact, Eq.~\ref{4dtopo} directly implies that
the 1-form symmetry charge (electric field) $\ve^2(x)$, which is
the variation $\delta S_{\mathrm{4d-topo}} / (\ii \delta
B^2_{01})$, equals to the flux of $B^1$, which is attached to the
monopole of $\va^1$.

The $3d$ boundary of the $4d$ SPT state is most naturally a
$(3+1)d$ QED$_4$ with both magnetic and electric 1-form
symmetries. The electric 1-form symmetry of the boundary QED$_4$
is inherited from the 1-form symmetry of $\va^1$ in the bulk,
while the magnetic 1-form symmetry of the QED$_4$ corresponds to
the electric 1-form symmetry of $\va^2$ in the bulk, because the
Dirac monopole line of $\va^1$ in the $4d$ bulk is bound/decorated
with the electric 1-form symmetry charge of $\va^2$. As we
mentioned previously, we will first discuss the situation with
bosonic point particles, hence in this QED$_4$ the infinitely
heavy electric charge and Dirac monopoles are both bosons. We
label this QED$_4$ as $\mathrm{QED}_4\{e_{0b}, m_{0b} \}$. Even
though these point particles have infinite mass, their statistics
still matter, because their Wilson loops (or 't Hooft loops) still
exist. If these point particles are fermions, the Wilson loop will
need a framing structure, and the Wilson loop or 't Hooft loop
with a twist will acquire a minus sign.

\subsection{Descendant $4d$ SPT state with $\U(1)^{(1)}\times Z_n^{(1)}$
symmetry}

Now we break one of the $\U(1)^{(1)}$ 1-form symmetry down to the
$Z_n^{(1)}$ symmetry. The topological response theory remains
unchanged from Eq.~\ref{4dtopo}, although one of the background
2-form gauge fields will become a $\mathcal{Z}_n^{(2)}$ background
2-form gauge field. The decorated monopole line construction
discussed in the previous section still applies here. One key
difference is that, because the $1d$ SPT phase with $Z_n^{(1)}$
1-form symmetry has a $\mathbb{Z}_n$ classification itself, the
flux attachment or decorated defect construction mentioned in the
previous subsection will naturally lead to a $\mathbb{Z}_n$
classification of the $4d$ SPT state also. Namely, when $k = n$ in
Eq.~\ref{4dtopo}, this bulk SPT state will be trivialized, because
the $1d$ SPT state decorated on the Dirac monopole line is
trivial.

We can always start with the QED$_4$ as a candidate boundary
state. Now since the magnetic 1-form symmetry is only $Z_n^{(1)}$,
it means that there are dynamical Dirac monopoles with
$n-$magnetic charges (Dirac monopole with $2\pi n$ flux quantum).
As we mentioned before we first focus on the cases where the point
excitations are bosons, then we can condense the $n-$magnetic
charge at the $3d$ boundary without breaking any symmetry. The
condensate of the $2\pi n$ Dirac monopole will drive the boundary
into a $3d$ $z_n$ topological order.

An ordinary $3d$ $z_n$ topological order is the deconfined phase
of a dynamical $z_n^{(1)}$ gauge field. In an ordinary $3d$ $z_n$
topological order, normally there are two types of excitations: a
point particle which is the remnant of the $2\pi$ Dirac monopole;
and also another line/loop excitation which is coupled to a
$z_n^{(2)}$ 2-form gauge field. If the loop excitation is
condensed (proliferated in $4D$ Euclidean space), the $z_n$
topological order is trivialized, and the system becomes gapped
and nondegenerate.

The dynamics of the loop excitation can be schematically described
by the following Hamiltonian \beqn H_{\mathrm{loop}} =
\sum_{\mathcal{C}} - t_{\mathcal{C}} \cos\left( \sum_{\vec{l} \in
\mathcal{C}} \hat{c}_{\vec{l}} - \sum_{\vec{p} \in
\mathcal{A}_\mathcal{C}} \hat{b}_{ \vec{p}} \right) + \cdots \eeqn
In this equation, $\mathcal{C}$ represents certain loop
configuration; $\vec{l}$ is a link which is part of this loop, and
$\mathcal{A}_\mathcal{C}$ is a membrane whose boundary is the loop
$\mathcal{C}$ ($\partial \mathcal{A_C} = \mathcal{C}$); $\vec{p}$
is a plaquette that belongs to $\mathcal{A_C}$.
$\Psi^\dagger_{\vec{l}} \sim \exp(\ii \hat{c}_{\vec{l}})$ is the
creation operator of the loop segment on link $\vec{l}$, and
$\hat{b}_{\vec{p}}$ is a 2-form gauge field defined on plaquette
$\vec{p}$. The direction of the link and the unit plaquette can be
absorbed into the definition of $\hat{c}$ and $\hat{b}$ and render
them a 1-form and 2-form fields.

For an ordinary $z_n$ topological order, both $\hat{c}_{\vec{l}}$
and $\hat{b}_{\vec{p}}$ take eigenvalues $2\pi N/n$ with integer
$N$. Hence the ``condensation" of the loop excitation will not
lead to degeneracy because of the existence of the $z_n^{(2)}$
2-form gauge field $\hat{b}$. Or in other words, the condensation
of the loop excitation will be fully ``Higgsed" due to the
coupling to the $z_n^{(2)}$ dynamical gauge field $\hat{b}$, and
this Higgs phase is the confined phase of the $z_n^{(1)}$ gauge
theory.

However, if the loop excitation carries a $\U(1)^{(1)}$ 1-form
charge, the situation would be very different. Now
$\hat{c}_{\vec{l}}$ can take continuous values between 0 and
$2\pi$. Condensing the loop would just drive the system back into
a gapless photon phase. Physically because the loop excitation
carries a $\U(1)^{(1)}$ 1-form charge, condensing the loop
excitations would lead to spontaneous $\U(1)^{(1)}$ 1-form
symmetry breaking, whose ``Goldstone mode" is precisely the
photon.

With the bulk response action Eq.~\ref{4dtopo}, the loop
excitation of $3d$ boundary carries charge quantum $k/n$ of the
$\U(1)^{(1)}$ 1-form symmetry. However, when $k = n$, the quantum
number of the loop excitation can be screened by binding with
unfractionalized integer 1-form symmetry charge, hence the loop
excitations become completely neutralized. Then when $k = n$ the
neutralized loop excitation can proliferate and drive the boundary
to a fully gapped and nondegenerate state, just like the case of
an ordinary $z_n^{(1)}$ gauge theory. This argument again leads to
a $\mathbb{Z}_n$ classification.

\subsection{Descendant $4d$ SPT state with $Z_q^{(1)}\times Z_n^{(1)}$
symmetry}

We can further break the left $\U(1)^{(1)}$ 1-form symmetry down
to $Z_q^{(1)}$ from the previous example. Now in the condensate of
the $2\pi n$ Dirac monopole, the loop excitation will carry $k/n$
unit of the $Z_q^{(1)}$ 1-form symmetry charge, and the loop
excitation is coupled to a dual $z_n^{(2)}$ gauge field. Our
interest is to ask when this $3d$ boundary can be fully gapped
without degeneracy.

Let us start with the simple example with $k = 1$, $q = 3$, and $n
= 2$. Following the discussion in the previous subsection, we
consider the $z_2$ topological order after condensing the $4\pi$
Dirac monopole at the boundary QED$_4$ (The $2\pi n$ monopole has
dynamics and can condense). There is a loop excitation of this
$z_2$ topological order, which couples to a dual $z_2^{(2)}$ gauge
field, and carries half charge of the $Z_3^{(1)}$ 1-form symmetry.
Now consider a loop excitation whose creation operator is
$P^\dagger_\mathcal{C}$: \beqn P^\dagger_\mathcal{C} \sim \prod_{
\vec{l} \in \mathcal{C}} \Psi^\dagger_{\vec{l}} \sim \exp(\ii
\sum_{\vec{l} \in \mathcal{C}} \hat{c}_{\vec{l}}). \eeqn
$P^\dagger_\mathcal{C}$ carries half charge under $Z_3^{(1)}$, and
it also couples to a dual $z_2^{(2)}$ gauge field. Under both the
$Z_3^{(1)}$ symmetry and the $z_2^{(2)}$ gauge symmetry,
$\mathcal{C}$ transforms as \beqn Z_3^{(1)}: P^\dagger_\mathcal{C}
\rightarrow e^{ \ii \frac{1}{2} \frac{2\pi N}{3} }
P^\dagger_\mathcal{C}, \cr \cr z_2^{(2)}-\mathrm{gauge}:
P^\dagger_\mathcal{C} \rightarrow - P^\dagger_\mathcal{C}, \eeqn
with integer $N$. One can check that by combining the loop
operator $P_\mathcal{C}$ with unfractionalized integer 1-form
charges, the $Z_3^{(1)}$ transformation can be completely
cancelled by a $z_2^{(2)}$ gauge transformation. In other words
the fractional $Z_3^{(1)}$ charge carried by the
$P^\dagger_\mathcal{C}$ can be ``neutralized" by binding a gauge
invariant $Z_3^{(1)}$ charge, and the $3d$ boundary system can be
driven into a trivial gapped phase by condensing this $Z_3^{(1)}$
neutral loop excitation.

The discussions above can be generalized to other $q$ and $n$.
With $k = 1 $ in Eq.~\ref{4dtopo}, after condensing the $2\pi n$
monopole, the $3d$ boundary system is driven into a $z_n$
topological order whose loop excitation carries $1/n$ fractional
$Z_q^{(1)}$ 1-form symmetry charge. Our interest is to check, when
this fractional 1-form symmetry charge can be ``neutralized" by
integer 1-form symmetry charge, namely by binding integer 1-form
symmetry charge the $Z_q^{(1)}$ transformation can be completely
absorbed/cancelled by the dual $z^{(2)}_n$ gauge transformation.

Under a $Z_q^{(1)}$ transformation, the loop creation operator
$P_\mathcal{C}$ acquires phase angle $2\pi/(nq)$; after binding
with $Q$ units of integer $Z_q^{(1)}$ charge, the loop would
acquire phase angle $2\pi/(nq) + 2\pi Q/q$. Now we seek for a pair
of integer $(Q, N)$ which suffices the following equation: \beqn
\frac{1}{nq} + \frac{Q}{q} = \frac{N}{n}. \eeqn This would mean
that the $Z_q^{(1)}$ transformation can be totally
absorbed/cancelled by a gauge transformation. For $(q,n) = (3,2)$
one can choose $(Q,N) = (1,1)$. In general the question is
equivalent to finding a pair of integers $(Q, N)$ that satisfies $
N q - Q n = 1$, which is only possible when $q$ and $n$ are
coprime. When $q$ and $n$ are not coprime, the loop quantum number
can be fully neutralized when $k = \mathrm{gcd}(q,n)$. This
implies a $\mathbb{Z}_{\mathrm{gcd}(q,n)}$ classification.

{\it --- More States}

All the SPT states discussed so far have bosonic electric charge
and Dirac monopoles at its boundary QED$_4$, namely the boundary
of all the SPT states are $\mathrm{QED}_4\{e_{0b}, m_{0b}\}$
states. Let us revisit the starting point of our bulk construction
of Eq.~\ref{4dtopo}. The two $u(1)$ gauge fields $\va^1$ and
$\va^2$ can have either bosonic or fermionic electric charges with
infinite mass in the bulk, which become the static electric
charges and Dirac monopoles of the boundary QED$_4$. Hence
logically there will also be $\mathrm{QED}_4\{e_{0b}, m_{0f}\}$,
$\mathrm{QED}_4\{e_{0f}, m_{0b}\}$, $\mathrm{QED}_4\{e_{0f},
m_{0f}\}$ states that we need to discuss. As we pointed out
before, the statistics of static particles still affect the
Wilson/'t Hooft loops. We defer discussions of these states to
section~\ref{4d3}.

\section{$4d$ SPT state with $\U(1)^{(1)} \times G$ symmetry and $3d$ Quantum Dimer model}

\label{4d2}

Here we consider $4d$ SPT states with both a $\U(1)^{(1)}$
symmetry and an ordinary 0-form symmetry $G$. The decorated defect
construction in the previous section can be generalized here: we
start with one $(4+1)d$ $u(1)$ gauge field $\va$ with a 1-form
electric symmetry, and decorate its Dirac monopole line with the
$1d$ SPT state with symmetry $G$, then condense the monopole line
in the bulk. A prototype $4d$ SPT state with such construction was
discussed previously, whose $G$ symmetry is $\SO(3)$, and its
topological response theory is~\cite{jian2020a} \beqn
\mathcal{S}_{\mathrm{4d-topo}} = \ii \pi \int_{(4+1)d} \
w_2[A^{\SO(3)}] \cup \frac{dB}{2\pi}, \label{4dtopo2} \eeqn where
$A^{\SO(3)}$ is the external 1-form $\SO(3)$ gauge field.

Generally speaking the discussion of $4d$ SPT state with 1-form
symmetry has implications on properties of $3d$ systems with
loop-like excitations. If in certain limit a $3d$ system with
spatial symmetries can be mapped to the boundary of a $4d$ state
with onsite symmetries, then whether or not the $4d$ bulk is a
nontrivial SPT state has strong implication on whether the $3d$
system can be trivially gapped or not, i.e. the nature of the $4d$
bulk helps us prove a Lieb-Schultz-Mattis (LSM)
theorem~\cite{LSM,hastings} of the $3d$ system. In recent years
much progress has been made in understanding the LSM theorems for
quantum spin systems using the anomaly analysis of its
corresponding higher dimensional bulk
states~\cite{Cheng_2016,xulsm,cholsm,maxlsm,chenglsm,masakilsm,elselsm}.
In condensed matter theories the quantum dimer model is an example
of systems with loop like excitations. Dimers are defined on the
links of the lattice, and each site of the lattice is connected to
a fixed number of dimers. Previous literature has shown that, the
$3d$ quantum dimer model can be mapped to a QED$_4$ without
dynamical electric charge~\cite{ms2003}, but its monopole can
carry nontrivial quantum number under spatial group due to the
Berry phase, and in particular, for the quantum dimer model on the
cubic lattice, the monopole of the QED$_4$ carries a ``spin-1/2"
representation (projective representation) of an emergent $\SO(3)$
symmetry~\cite{senthilloop,balentsdimer}. Hence this quantum dimer
model is analogous to the boundary of a $4d$ SPT state with
symmetry $\U(1)^{(1)} \times \SO(3)$, and there should be a LSM
theorem for this quantum dimer model.

This LSM theorem for the quantum dimer model is consistent with
the LSM theorem for spin-1/2 systems on the cubic lattice. In
Ref.~\onlinecite{xulsm}, various quantum spin systems on the cubic
lattice were considered. For example, a $\SU(N)$ spin system on
the cubic lattice with fundamental and antifundamental
representations on the two sublattices of the cubic lattice has a
LSM theorem for even integer $N$, but there is no LSM theorem for
odd integer $N$, i.e. the quantum spin system described above with
odd integer $N$ can have a featureless gapped ground state on the
cubic lattice. However, a quantum dimer model on the cubic lattice
could be the low energy effective description of all these
systems, since two nearest neighbor AB sites can always form a
dimer (spin singlet), regardless of even or odd integer $N$.

One simple extension of Eq.~\ref{4dtopo2} is that, when we break
$\SO(3)$ down to its subgroup $\U(1) \rtimes Z_2$,
Eq.~\ref{4dtopo2} reduces to \beqn S_{\mathrm{4d-topo}} = \ii
\frac{\Theta}{(2\pi)^2} \int_{(4+1)d} \ dB \wedge dA,
\label{4dtopo3} \eeqn where $A$ is the background $\U(1)$ gauge
field. The integral in Eq.~\ref{4dtopo3} is quantized, hence
$\Theta$ is periodically defined: $\Theta = \Theta + 2\pi$. Under
the $Z_2$ subgroup of $\SO(3)$, $A$ changes sign, hence a
symmetric response theory demands $\Theta = k \pi$ with integer
$k$. Eq.~\ref{4dtopo3} with $k = 1$ corresponds to the nontrivial
$4d$ SPT phase.

Eq.~\ref{4dtopo3} also describes the corresponding $4d$ bulk state
if instead we consider a quantum dimer model defined on a $3d$
tetragonal lattice, here the $\U(1)$ symmetry is further reduced
to a $Z_4$ symmetry, and the $Z_4$ corresponds to the rotation of
the square lattice in each layer. In this case in the topological
response theory Eq.~\ref{4dtopo3}, $A$ is a background $Z_4$ gauge
field. Eq.~\ref{4dtopo3} still describes a nontrivial $4d$ SPT
state with 1-form symmetry.


The situation will be very different if we consider a quantum
dimer model on a $3d$ bipartite lattice with an effective $Z_3
\rtimes Z_2 = S^3$ symmetry.
The $Z_3$ should correspond to a three fold rotation $C_3$ in the
XY plane, and $Z_2$ is a $\pi$-rotation about the $x$-axis. Such
quantum dimer models can potentially be mapped to the boundary of
a $4d$ system with $\U(1)^{(1)} \times S^3$ symmetry. But there is
no $1d$ SPT state with the $S^3$ symmetry, hence the $4d$ bulk
with the $\U(1)^{(1)} \times S^3$ symmetry is also trivial as a
descendant state of the SPT state described by Eq.~\ref{4dtopo3}.
Hence there should be no LSM theorem for these quantum dimer
models, i.e. these quantum dimer models can in general have a
gapped ground state without degeneracy, unless this model has
higher symmetries than the lattice itself.

\section{Other $4d$ SPT states}

\label{4d3}

With just a $\U(1)^{(1)}$ symmetry, there is already a nontrivial
$4d$ SPT phase, whose boundary is a QED$_4$ with a 1-form electric
symmetry, and the Dirac monopole is a fermion (labelled as $m_f$).
The unit electric charge (labelled as $e_{0b}$) is infinitely
heavy at the boundary QED$_4$ due to the $\U(1)^{(1)}$ symmetry.
We label this boundary QED$_4$ as state $\mathrm{QED}_4\{e_{0b},
m_f\}$. The bulk is a nontrivial SPT state, namely its boundary
QED$_4$ cannot be trivially gapped. One can condense a Cooper pair
of the fermionic Dirac monopole $m_f$, and drive the QED$_4$ to a
``monopole superconductor", which is also a $z_2$ topological
order. The loop excitation of the $z_2$ topological order will
carry a fractional half charge of the $\U(1)^{(1)}$ 1-form
symmetry, and hence cannot lead to a fully gapped and
nondegenerate state after condensation for the reasons explained
previously in this manuscript. Although the electric charges are
infinitely heavy due to the 1-form symmetry, its statistics still
matters to physical observables such as the Wilson loops of the
QED$_4$. And in this QED$_4$ the infinitely heavy electric charge
is a boson.

This state remains a nontrivial SPT after breaking the
$\U(1)^{(1)}$ down to $Z_n^{(1)}$ with even integer $n$, the cases
with $n = 2, 4$ were discussed in Ref.~\onlinecite{juven,juven2}.
But this state will be trivialized if $n$ is an odd integer. For
odd integer $n$, in the monopole superconductor constructed above,
the loop excitation carries half charge of the $Z_n^{(1)}$ 1-form
symmetry, and it can be ``neutralized" by binding unfractionalized
1-form symmetry charge, i.e. the $Z_n^{(1)}$ transformation on the
loop excitation can be completely cancelled by the $z_2^{(2)}$
gauge transformation on the loop excitation, then the condensation
of the neutralized loop can lead to a trivially gapped phase.

There is even a nontrivial bosonic SPT state in $4d$ space without
any symmetry; its boundary is a QED$_4$ whose both electric charge
and Dirac monopole (including their bound state dyon) are
fermions~\cite{wangpottersenthil,allfermion}. We label this QED as
$\mathrm{QED}_4\{e'_f, m'_f\}$ state. We view the
$\mathrm{QED}_4\{e_{0b}, m_f\}$ and $\mathrm{QED}_4\{e'_f, m'_f\}$
as two root states, and by ``gluing" these two QED$_4$ states
together, another new state can be constructed. One can condense
the bound state of the Dirac monopoles (labelled as $(m_f, m'_f)$)
of both QED$_4$ systems, then the gauge fields from both QED$_4$
will be identified due to the Higgs mechanism, and $e_{0b}$ and
$e_f$ are both confined since they both have nontrivial statistics
with the condensed bound state of monopoles. Although $e_{0b}$ is
infinitely heavy, its confinement can still be defined by the
behavior of Wilson loop of its gauge field. In the condensed phase
of bound state $(m_f, m'_f)$, the Wilson loop of each individual
gauge field obeys the area law. But the bound state $(e_{0b}, -
e'_f)$, which has trivial mutual statistics with $(m_f,m'_f)$,
remains deconfined, though it is still infinitely heavy. This new
QED state has infinitely heavy fermionic electric charge, and
dynamical fermionic Dirac monopole. This new state is labelled as
$\mathrm{QED}_4\{e_{0f}, m_f\}$. One can also exchange $e$ and
$m$, and label the state as $\mathrm{QED}_4\{e_{f}, m_{0f}\}$,
i.e. a state with dynamical fermionic gauge charge, but infinitely
heavy fermionic Dirac monopole.

$ $

{\it \underline{Summary of $4d$ SPT states with 1-form
symmetries:}}

Let us reinvestigate the states discussed in the end of
section~\ref{4d1}. As we briefly discussed there, besides the
states $\mathrm{QED}_4\{e_{0b}, m_{0b}\}$, logically there should
also be $\mathrm{QED}_4\{e_{0b}, m_{0f}\}$,
$\mathrm{QED}_4\{e_{0f}, m_{0b}\}$, $\mathrm{QED}_4\{e_{0f},
m_{0f}\}$, which can all be boundary states of $(4+1)d$ SPT bulk.
It turns out that these states can be constructed by gluing states
in section~\ref{4d1} and \ref{4d3}. For example, starting with the
state $\mathrm{QED}_4\{e_{0b}, m_{0b}\}$ discussed in
section~\ref{4d1} (we label its gauge field as $\vec{a}$), one can
combine it with the state $\mathrm{QED}_4\{e'_{0b}, m'_{f}\}$
(with gauge field $\vec{a}'$) discussed in section~\ref{4d3}, and
consider the charge bound state $(e_{0b}, -e'_{0b})$. This bound
state carries zero total gauge charge of $\vec{a}$ and $\vec{a}'$.
We assume that there is only one $\U(1)^{(1)}$ 1-form symmetry,
hence the charge bound state $(e_{0b}, -e'_{0b})$, which carries
zero total gauge charge, is no longer necessarily infinitely heavy
and can acquire dynamics and condense. Its condensate would render
$\vec{a} = \vec{a}'$ through the Higgs mechanism, and in the
condensate the monopole bound state $(m_{0b}, m'_f)$ remains
deconfined, as it has trivial mutual statistics with $(e_{0b},
-e'_{0b})$. The final state is identical to state
$\mathrm{QED}_4\{e_{0b}, m_{0f}\}$ discussed in section~\ref{4d1}.
Following the same argument, through gluing
$\mathrm{QED}_4\{e_{0b}, m_{0f}\}$ and state
$\mathrm{QED}_4\{e_{f}', m_{0f}'\}$ discussed in section~\ref{4d3}
(by condensing the bound state $(m_{0f}, - m'_{0f})$), one can
obtain another state $\mathrm{QED}_4\{e_{0f}, m_{0f}\}$ discussed
in section~\ref{4d1}.

The construction of all these states discussed so far can be
summarized mathematically in a single unified topological response
theory in the $(4+1)d$ bulk: \beqn && S_{\mathrm{4d-topo}} =
\int_{(4+1)d} \ \frac{\ii k_0}{2\pi} B^1 \wedge d B^2 \cr\cr &+&
\frac{\ii k_1}{2} dB^1 \cup w_2 + \frac{\ii k_2}{2} dB^2 \cup w_2
+ \ii \pi k_3 w_2 \cup w_3. \label{4dtopofull} \eeqn $w_2$ and
$w_3$ are the second and third Stiefel-Whitney class of the
space-time manifold. $k_0$ takes arbitrary integer values, while
$k_1$, $k_2$ and $k_3$ only take value 0 and 1, since the
Stiefel-Whitney class is defined mod 2. This topological response
theory is equivalent to the discussion based on the cobordism
theory in Ref.~\onlinecite{juven,juven2}.

The classification of $4d$ SPT states discussed so far is
summarized as follows:
 \beqn
 \U(1)^{(1)} &:& \mathbb{Z}_2 \otimes \mathbb{Z}_2;
 \cr\cr
 Z_n^{(1)} &:& \mathbb{Z}_{2} \otimes \mathbb{Z}_{\mathrm{gcd}(2,n)};
 \cr\cr
 \U(1)^{(1)} \times \U(1)^{(1)} &:& \mathbb{Z} \otimes
 \mathbb{Z}_{2}^3;
 \cr\cr
 \U(1)^{(1)} \times Z_n^{(1)} &:& \mathbb{Z}_n \otimes \mathbb{Z}_{2}^2
 \otimes \mathbb{Z}_{\mathrm{gcd}(2,n)};
 \cr\cr
 Z_q^{(1)} \times Z_n^{(1)} &:& \mathbb{Z}_{\mathrm{gcd}(q,n)} \otimes \mathbb{Z}_{\mathrm{gcd}(2,q)}
 \otimes \cr\cr &&
 \mathbb{Z}_{\mathrm{gcd}(2,n)} \otimes
 \mathbb{Z}_{2} .
 \eeqn

\section{$3d$ SPT state with $G_1^{(1)}\times
G_2$ symmetry}

\subsection{Parent $3d$ SPT state with $\U(1)^{(1)} \times \U(1)$ symmetry}

The parent $3d$ SPT state we will consider, is a state with
$\U(1)^{(1)}\times \U(1)$ symmetry. We can couple its symmetry
currents to a background 2-form gauge field $B$, and a 1-form
gauge field $A$. The response theory for this SPT state is \beqn
S_{\mathrm{3d-topo}} = \int \ \frac{\ii k}{2\pi} B \wedge dA =
\int \ \frac{\ii k}{2\pi} A \wedge dB. \label{3dtopo} \eeqn To
construct such state, again one can rely on the decorated defect
picture. We can start with a photon phase with an electric
$\U(1)^{(1)}$ 1-form symmetry, namely there is no dynamical
electric charge, or equivalently the electric charge is infinitely
heavy, but there are dynamical Dirac monopoles. Then we decorate
the Dirac monopole with a zero dimensional bosonic SPT state with
$\U(1)$ symmetry, which is a bosonic charge with $\U(1)$ symmetry.
This zero dimensional bosonic SPT state has $\mathbb{Z}$
classification, which correspond to states with integer charges of
a boson with $\U(1)$ symmetry. These states can also be
equivalently constructed by decorating the vortex line of the
$\U(1)$ order parameter with a $1d$ SPT state with $\U(1)^{(1)}$
1-form symmetry, i.e. the building bricks discussed in
section~\ref{1d}.

After condensing the decorated Dirac monopole, the $3d$ bulk of
the system is driven into a fully gapped state without degeneracy.
The $2d$ boundary of the system would most naturally be a QED$_3$
whose dynamical $u(1)$ gauge field $\va$ has no dynamical gauge
charge, but its magnetic flux carries conserved $\U(1)$ quantum
number that couples to $A$. The QED$_3$ is a dual of the
superfluid phase with spontaneous breaking of the $\U(1)$
symmetry. And the assumption that there is no dynamical electric
charge of gauge field $\va$ is equivalent to the statement that
there is no dynamical vortex of the dual superfluid, hence the
superfluid cannot be disordered by condensing the vortices.

\subsection{Descendant $3d$ SPT state with $\U(1)^{(1)} \times Z_n$ symmetry}

We can break the $\U(1)$ 0-form symmetry coupled to $A$ in
Eq.~\ref{3dtopo} down to a $Z_n$ symmetry, now the entire symmetry
becomes $\U(1)^{(1)}\times Z_n$. The topological response theory
Eq.~\ref{3dtopo} still applies, but now $A$ becomes a
$\mathcal{Z}^{(1)}_n$ background gauge field. The decorated defect
construction in the previous case would lead to a $\mathbb{Z}_n$
classification, because the zero dimensional SPT state with $Z_n$
symmetry decorated at the Dirac monopole has a $\mathbb{Z}_n$
classification.

This classification can be understood at the boundary as well. The
$(2+1)d$ boundary is a QED$_3$ whose flux carries $k$ units of the
$Z_n$ quantum number, where $k$ is given in Eq.~\ref{3dtopo}. With
$k = n$, the flux of the QED$_3$ basically carries trivial quantum
number, and the QED$_3$ can be driven into a trivial confined
phase. This boundary state is similar to the quantum dimer model
on a $2d$ bipartite lattice, such as the square lattice. The
quantum dimer model can be mapped to a compact QED$_3$ with no
electric charge (the quantum dimer constraint, i.e. every site is
connected to precisely one dimer, is strictly enforced), but the
flux of the compact QED$_3$ carries nontrivial lattice quantum
number. The description of the quantum dimer model in terms of
QED$_3$ is analogous to the boundary of the $3d$ SPT state with
$\U(1)^{(1)} \times Z_4$ symmetry at $k=1$. It is well-known that
the confined phase of the quantum dimer model on the square
lattice cannot be a trivial gapped phase, instead it must have
ground degeneracy due to spontaneous breaking of lattice symmetry.
But in the quantum dimer model because the $Z_4$ symmetry is a
non-onsite lattice symmetry, the quantum dimer model exists as a
well defined system in $2d$.

This effect is inherited from the LSM theorem for spin-1/2 systems
on the square lattice. There is no LSM theorem for a spin-2 system
on the square lattice, and a spin-2 system can be viewed as four
copies of spin-1/2 systems glued together, or a system with four
spin-1/2s in each unit cell. All these observations are consistent
with the $\mathbb{Z}_4$ classification of the $3d$ SPT state with
$\U(1)^{(1)} \times Z_4$ symmetry discussed in this section.

\subsection{Descendant $3d$ SPT state with $Z_q^{(1)} \times \U(1)$ symmetry}

Next we consider the $3d$ SPT states as descendant states of
Eq.~\ref{3dtopo} with $Z_q^{(1)}\times \U(1)$ symmetry. Again we
will first consider the cases where all the point particles in the
bulk are bosons. When we break the $\U(1)^{(1)}$ symmetry down to
$Z_q^{(1)}$, the $2d$ boundary is a QED$_3$ whose flux carries
$\U(1)$ quantum number, and there are dynamical $q-$fold electric
charges. The boundary can only be driven to a $z_q$ topological
order by condensing the $q-$fold electric charge. One of the point
like anyons of this topological order is the remnant of the
$2\pi/q$ flux of the QED$_3$, which carries $k/q$ charges of the
$\U(1)$ symmetry quantum number. When $k = q$ this anyon carries
unfractionalized quantum number, hence can be neutralized by
binding with gauge invariant integer charge of the $\U(1)$
symmetry. This neutralized anyon is a self-boson, and after
condensation it drives the boundary into a trivial gapped state.
Hence this $3d$ SPT state should have a $\mathbb{Z}_q$
classification.

To facilitate further discussions let us also consider a different
$3d$ bulk state with $\U(1)$ global symmetry only. This is a
QED$_4$ whose electric charge is fermion, and Dirac monopole is a
boson (using the notations introduced before, this bulk state is
$\mathrm{QED}_4\{ e_f, m_b \}$). Again one can bind the Dirac
monopole with another boson that carries $\U(1)$ quantum number,
and condense the bound state in the $3d$ bulk. Then the bulk is
gapped and nondegenerate, while the $2d$ boundary is a QED$_3$
whose electric charge is a fermion, while the gauge flux carries
$\U(1)$ quantum number. However, this $3d$ bulk is not a SPT
state, since one can put the electric charge at the boundary in a
$2d$ Chern insulator with Hall conductivity 1, then the $2d$
boundary is gapped without breaking any symmetry. This is
consistent with the classification of ordinary SPT states without
higher form symmetries. With only $\U(1)$ symmetry, there is no
nontrivial SPT state in $3d$. One needs another time-reversal
symmetry to construct a $3d$ bosonic SPT state, since the boundary
Chern insulator of the fermionic gauge charge as we constructed
above necessarily breaks the time-reversal.

One can again glue the $2d$ boundary states in the previous two
paragraphs together. Let us recall that the boundary of a
nontrivial $3d$ SPT state with $Z_q^{(1)} \times \U(1)$ symmetry
is a QED$_3$ whose flux carries $\U(1)$ quantum number, and its
bosonic electric charges are infinitely heavy; the boundary of the
trivial state discussed in the last paragraph is a QED$_3$ whose
flux also carries $\U(1)$ quantum number, and its electric charge
is a fermion with nonzero dynamics. Once we couple the two $2d$
systems together, the tunnelling between the gauge fluxes between
the two QED$_3$ will be turned on, which identifies the two gauge
fields. Now the $2d$ boundary state is a QED$_3$ whose gauge flux
still carries $\U(1)$ quantum number, but its static electric
charge is a fermion. This state is not a new SPT state since it
can be constructed by gluing the $2d$ boundaries of the two
systems discussed above.

\subsection{Descendant $3d$ SPT state with $Z_q^{(1)}\times
Z_n$ symmetry}

Finally we can break the $\U(1)^{(1)}$ 1-form symmetry in
Eq.~\ref{3dtopo} to $Z_q^{(1)}$. Again we can start with the
QED$_3$ state at the $(2+1)d$ boundary. In this case there are
dynamical $q$-fold electric charge of the $u(1)$ gauge field, and
the magnetic flux of the $u(1)$ gauge field still carries $Z_n$
quantum number. One can condense the charge$-q$ bound state, and
drive the $2d$ boundary into a $2d$ $z_q$ topological order. In an
ordinary $2d$ $z_q$ topological order, there are two sets of
anyons. The $e$ anyon is a remnant of the unit charge excitation
of the QED$_3$ before the condensation of the $q$-fold electric
charge, and the $m$ anyon is a $2\pi/q$ flux quantum of the $u(1)$
gauge flux. Both $e$ and $m$ anyons are self-bosons, but have a
mutual $2\pi/q$ statistical angle. In our current case, due to the
$Z_q^{(1)}$ 1-form symmetry, the $e$ anyons are not dynamical, and
a $m$ anyon carries a fractional quantum number $1/q$ of the $Z_n$
symmetry (assuming $k=1$ in Eq.~\ref{3dtopo}). Both $e$ and $m$
anyons are coupled to $z_q$ gauge fields. Following the arguments
in section~\ref{4d1}, we can demonstrate that when $q$ and $n$ are
coprime, the fractional quantum number of the $m$ anyon can always
be ``neutralized" by binding with integer charges of the $Z_n$
symmetry, in the sense that the $Z_n$ transformation on the
decorated $m$ anyon can always be cancelled by a $z_q$ gauge
transformation. When $q$ and $n$ are not coprime, the quantum
number of the $m$ anyon can be neutralized when $k =
\mathrm{gcd}(q,n)$. The neutralized $m$ anyon can condense and
drive the $2d$ boundary to a trivial gapped state without
degeneracy. Hence as a descendant state of Eq.~\ref{3dtopo}, the
classification of the $3d$ SPT state with $Z_q^{(1)}\times Z_n$
symmetry is $\mathbb{Z}_{\mathrm{gcd}(q,n)}$.

$ $

{\it \underline{Summary of $3d$ SPT states with 1-form
symmetries:}}

Here we summarize the classification of $3d$ SPT states that are
descendants of Eq.~\ref{3dtopo}. If there are special SPT states
that cannot be described by Eq.~\ref{3dtopo}, such as some of the
states discussed in Ref.~\onlinecite{wen1formspt1,wen1formspt2},
they are not included in this list. \beqn
 \U(1)^{(1)} \times \U(1) &:& \mathbb{Z};
 \cr\cr
 Z_q^{(1)} \times \U(1) &:& \mathbb{Z}_q;
 \cr\cr
 \U(1)^{(1)} \times Z_n &:& \mathbb{Z}_n;
 \cr\cr
 Z_q^{(1)} \times Z_n &:& \mathbb{Z}_{\mathrm{gcd}(q,n)}.
 \eeqn

\section{$2d$ SPT state with $G_1^{(1)}\times
Z_2^T$ symmetry}

Several different $(2+1)d$ SPT states that involve 1-form
symmetries can be described by the following topological response
term: \beqn S_{\mathrm{2d-topo}} = \int_{(2+1)d} \ \frac{\ii
\Theta}{2\pi}dB \label{2dtopo} \eeqn In principle $\Theta$ can
take arbitrary value, because $dB$ is gauge invariant. But some
extra symmetry can pin $\Theta$ to a specific value, like the
$\Theta$ term of the ordinary topological insulator~\cite{qi2008}
and bosonic SPT state~\cite{senthilashvin}.

As an example of such states, we assume that the 2-form background
gauge field $B$ is unchanged under time-reversal transformation,
this means that the 1-form symmetry charge will change sign under
time-reversal. This implies that the total symmetry of the system
is a direct product between the 1-form symmetry and time-reversal.
$\Theta$ is clearly defined periodically, namely $\Theta + 2\pi =
\Theta$, hence the time-reversal invariant states correspond to
$\Theta = \pi k$ with arbitrary integer $k$.

For even integer $k$, the $(2+1)d$ topological response theory
Eq.~\ref{2dtopo} reduces to a boundary topological term that is
identical to the topological response theory with $1d$ SPT state
with a 1-form symmetry (section~\ref{1d}). This means that, for
even integer $k$, the boundary corresponds to a well-defined $1d$
state, hence an even integer $k$ would correspond to a trivial
state in $(2+1)d$. On the other hand, for odd integer $k$, the
boundary is a ``half" $1d$ SPT state with 1-form symmetry
$G^{(1)}$. Then the $(2+1)d$ bulk could be a SPT state.

As we mentioned before, due to the strict constraint $\nabla_x
\hat{e}(x)  = 0$ for 1-form charge in one dimension, a $1d$ system
with 1-form symmetry is analogous to a $0d$ system with ordinary
0-form symmetry. Then whether there is a $(2+1)d$ SPT state with
$G^{(1)} \times Z_2^T$ symmetry can also be determined by the
existence of projective representation of $G \times Z_2^T$. And
there is a 2-dimensional projective representation of $\U(1)
\times Z_2^T$, but not for $\U(1) \rtimes Z_2^T$. Indeed, if the
symmetry of the system is $G^{(1)} \rtimes Z_2^T$, namely $B$ is
odd under time-reversal, the $\Theta$ coefficient is unchanged
under time-reversal, hence time-reversal will not pin $\Theta$ to
any specific value.

To summarize our result in two spatial dimensions, there is a
nontrivial $2d$ SPT state with $\U(1)^{(1)} \times Z_2^T$
symmetry, and this state remains nontrivial when $\U(1)^{(1)}$ is
broken down to $Z_q^{(1)}$ with even integer $q$.

The decorated defect construction also applies in this scenario,
which is analogous to what was discussed in
Ref.~\onlinecite{chenluashvin} for ordinary SPT states. We can
construct the SPT state with $k = 1$ in Eq.~\ref{2dtopo}, by first
creating a domain wall of time-reversal symmetry, then embed each
domain wall with a $1d$ SPT state described by Eq.~\ref{1dtopo},
and finally proliferate the domain walls. Besides construction
from $1d$ SPT state, we can also obtain this $2d$ SPT state by
reduction from higher dimensions. For example, starting with the
$3d$ SPT state with $\U(1)^{(1)} \times \U(1)$ symmetry described
by the response theory Eq.~\ref{3dtopo}, one can compactify one of
the three spatial dimensions (the $3d$ space $\mathrm{R}^3$
becomes $\mathrm{R}^2 \otimes S^1$), and insert a $\pi-$flux of
the 1-form gauge field $A$ through $S^1$. Then the response theory
Eq.~\ref{3dtopo} reduces to Eq.~\ref{2dtopo} with $k = 1$. This is
the same procedure of dimensional reduction introduced in
Ref.~\onlinecite{qi2008}.

\section{Discussion}

In this work we discussed the classification, construction, and
boundary properties of SPT states involving higher symmetries,
from one to four spatial dimensions. Our discussion is mostly
based on physical arguments. As an application of our discussion,
we make connection between the SPT states with 1-form symmetry to
quantum dimer model at one lower dimension. Quantum dimer model
with spatial symmetries can be mapped to the boundary of a bulk
state with onsite symmetries. Some of the universal features of
the quantum dimer model is dictated by the nature of the
corresponding bulk state.

In this work we only discussed quantum dimer models on bipartite
lattices, which can be mapped to a QED with $\U(1)^{(1)}$ 1-form
symmetry. It is well known that some other dimer models can be
naturally mapped to a $z_2$ gauge field, such as quantum dimer
model on the triangular lattice~\cite{ms2001}. Then these models
would be examples of systems with $Z_2^{(1)}$ 1-form symmetry, and
they can also be potentially mapped to the boundary of one higher
dimensions. Insights for these systems gained from higher
dimensions will be studied in later works.

This work is supported by NSF Grant No. DMR-1920434, the David and
Lucile Packard Foundation, and the Simons Foundation. The authors
thank Wenjie Ji for very helpful discussions.

\bibliography{form}

\end{document}